\documentclass[doublecol,linenumbers]{epl2} 

\usepackage[english]{babel}
\usepackage[utf8]{inputenc}
\usepackage{graphicx}

\title{Rich or poor: who should pay higher tax rates?}

\author{Paulo Murilo Castro de Oliveira \inst{1,2}}

\institute{                    
  \inst{1} Universidade Federal da Integração Latino Americana - Foz do Iguaçu PR, Brasil\\
  \inst{2} Instituto de Física, Universidade Federal Fluminense - Niterói RJ, Brasil\\
  pmco@if.uff.br, oliveira.paulomurilo@gmail.com
}


\abstract{
 A dynamic agent model is introduced with an annual random wealth multiplicative process
 followed by taxes paid according to a linear wealth-dependent tax rate. If poor agents pay higher
 tax rates than rich agents, eventually all wealth becomes concentrated in the hands 
 of a single agent. By contrast, if poor agents are subject to lower tax rates, the economic collective
 process continues forever. 
}

\begin{document}

\maketitle

\section{Introduction}

	Inequality within human societies is nowadays, more than ever, an important issue 
 \cite{Science}. Historical data were collected and analysed in a recent, now famous study 
 \cite{Piketty}, concerning individuals, regions, countries, etc. Within this general frame, 
 progressive taxation (rich people paying higher tax rates) is proposed in order to mitigate 
 the observed tendency towards wealth concentration. On the other hand, some argue the contrary, 
 rich people paying less taxes, because somehow their wealth benefits the whole society 
 (the trickle-down argument). Here, we adopt the strategy of summing up the whole wealth of a  
 population formed by a set of agents $n = 1, 2 \dots N$, and studying the time evolution of 
 the share $w_n$ each individual owns of this total wealth. A simple model is studied via 
 numerical simulation as well as via an analytic, mean-field approach. The model is based 
 on two general ingredients, annual profits and taxation. We focus attention on the final 
 steady state distribution of wealth. If the share of some particular agent approaches 
 the maximum possible value $w_{\rm max} = 1$ (this  possibility is called {\sl collapse}), 
 the wealth of the entire population lies in hands of a single agent. 
 Economic evolution stops, as this is an absorbing state of the dynamics. This kind of dynamic 
 transition towards possible absorbing states is a recent research field \cite{MarroDickman}, 
 where models as the present one are used in order to study such phenomena as extinction, 
 epidemics, opinion dynamics, vaccination, prevention of fires, etc. In the present 
 case, the general strategy of following the wealth shares (instead of wealth values themselves) 
 distinguishes this study from similar models. This subject was reviewed in \cite{Yakovenko}, 
 describing the different strategies adopted in these models in order to represent the economic 
 dynamics of a society. The simplest one is the pairwise transaction, where two randomly chosen 
 agents exchange money according to some conservative rule. The inclusion of some external entity 
 (a bank or some reservoir of money) allows the inclusion of debts, the total money of the 
 population is no longer conserved. The concept of wealth instead of money considers other 
 individual property, besides money. The total population wealth is not conserved, since the 
 annual production of each agent can increase (or decrease) its wealth. This ingredient is 
 modeled by multiplying the current wealth of each agent by a random factor in the pioneering 
 work of Bouchaud and M\'ezard \cite{Bouchaud}. We provide in Ref. \cite{many} an incomplete and 
 somewhat arbitrary list of works related to this issue; the interested reader may also wish to 
 consult the works cited in these references. All of them follow the dynamic evolution of the 
 wealth distribution among agents, not the distribution of wealth shares here introduced. Concerning 
 taxes, they were treated in \cite{Bouchaud} as uniformly applied to each agent independent of its 
 current wealth, the possibility of regressive or progressive tax rates is also introduced here.

\section{Stochastic, Computer Simulated Version}

	Consider a population of $N$ independent agents, each one owning a positive wealth $W_n$\footnote{
 The set of wealths is always considered in decreasing order, the so-called Zipf distribution, so index $n$ 
 corresponds to the rank of each agent, not to the specific agent itself.}. The total 
 population wealth is $S = \sum_n W_n$. The share of this total wealth owned by agent $n$ is $w_n = W_n/S$. 
 Consider also the monetary unit taken as the total wealth $S = 1$. Using this unit, wealths $W_n$ or wealth 
 shares $w_n$ are the same quantities.

	The dynamic evolution consists of four steps during each ``year'' (time step). Step I is a 
 multiplicative process with randomness. Each wealth $w_n$ is multiplied by a random, positive 
 factor $f_n$ chosen from a fixed probability distribution. The result is a new set of $W'_n = f_n w_n$ 
 and the corresponding shares $w'_n = W'_n/S'$, where $S' = \sum_n W'_n$. Second, step II, payment of 
 taxes at the end-of-year, according to a tax rate that is a linear function of wealth,
 $W''_n = (1-A-p\,w'_n)\,W'_n$, where $A$ and $p$ are fixed parameters obeying the restrictions $0<A<1$ and 
 $0<A+p<1$. Step III is a partial redistribution procedure: a fraction $R$ of the total collected taxes is 
 uniformly redistributed among all $N$ agents, $W'''_n = W''_n + R\,\sum_n (A+p\,w'_n)\,W'_n/N$. Numerically, 
 it provides the advantage of restricting from below the dynamic variables $w_n$, so that they are 
 always strictly positive. This restriction is mandatory, otherwise agents reaching $w_n=0$ would be removed 
 from the game permanently. The interesting 
 limiting case, however, is $R \to 0$, as we shall see. Finally, step IV is the renormalisation of 
 wealths by a common factor, $w'''_n = W'''_n/S'''$, where $S''' = \sum_n W'''_n$. This step 
 can be interpreted as a simple redefinition of the monetary unit, always kept equal to the total 
 wealth at the beginning of the next year. After this four-step procedure, the new year starts 
 from the set of $w'''_n$, and so on. Besides the focus on wealth shares instead of wealths, another 
 difference of the current model compared with \cite{Bouchaud} is the presence of the non-linear (quadratic 
 in $W_n$) term in the tax paid (for $p \ne 0$).

	Anticipating the result, in the quoted limit of no-redistribution, $R \to 0$, there is an absorbing 
 state transition between two possible final steady states, depending on the adopted $W$-dependent tax 
 rate rule. Collapsed phase occurs when poor agents pay higher tax rates than rich agents, $p<0$, 
 resulting in one particular agent owning all the wealth at the end, 
 $w_1 = w_{\rm max} \to 1$ for the richest agent, $w_n \to 0$ for all others. The economy ceases to evolve. 
 Otherwise, an active phase occurs when rich agents pay higher tax rates than poor agents, $p>0$, 
 the whole economy evolves forever with the wealth distributed among agents. Economic evolution 
 survives. Being a simple global rescaling, parameter $A$ plays no essential role.
 It is used only to assure all tax rates are positive. Also, the same model with 
 taxes applied only to the annual gains (instead of accumulated wealths) exhibits the same transition.

\begin{figure}
\onefigure[width=\columnwidth]{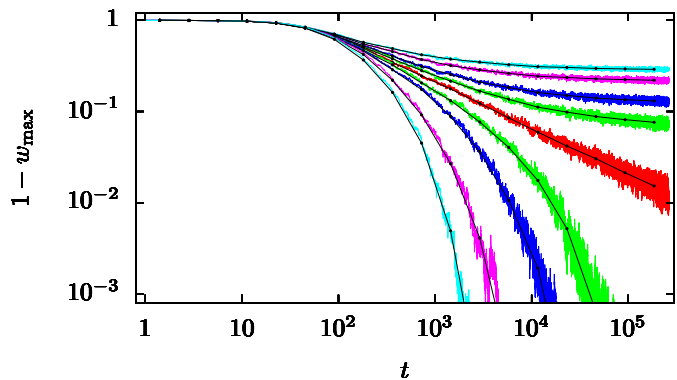}
 \caption{Sum of all wealth shares but the richest agent as a function of time. The central red curve 
  corresponds to $p=0$, the critical situation. Black points show the averages in powers-of-2 
  increasing intervals (bins) of the same data, exhibiting a clear long-term linear behaviour in this 
  $log \times log$ plot, thus a power law time dependence. Out from the central curve, 
  green curves correspond to $p=\pm 0.005$, blue curves to $p=\pm 0.01$, purple to $p=\pm 0.02$ and 
  light blue to $p=\pm 0.03$.}
\label{fig.1}
\end{figure}

\begin{figure}
\onefigure[width=\columnwidth]{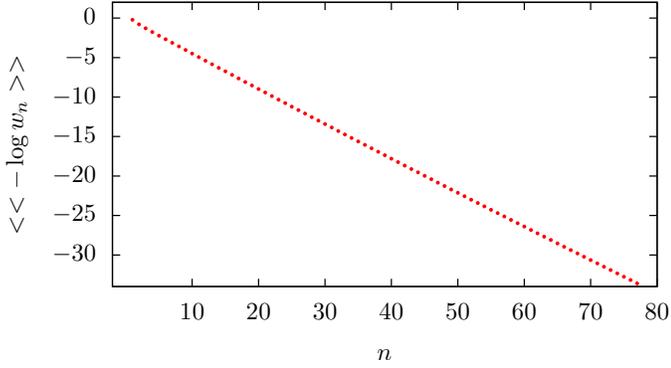}
\caption{Steady state distribution of wealth shares as a function of the rank 
 $n = 1, 2 \dots N$ (logarithm base $10$). Error bars are smaller than the symbols.
 Here, $N=1,000$ and $p=0.045$.}
\label{fig.2}
\end{figure}

	The same scenario, transition to a collapsed state for $p < 0$, is observed with 
 numerous alternative probability distributions for the multiplicative factors $f_n$ adopted in step I. A 
 simple possibility is to double the current wealth $W_n$ ($f_n = 2$) with probability $50\%$, and otherwise 
 leave it unaltered ($f_n = 1$). Figure 1 shows the result for this simple choice, in a population of 
 $N = 1,000$ agents. Many other rules for selecting the multiplicative factors $f_n$ were tested, all of them 
 giving the same general result, a dynamically-induced transition at the critical point 
 $p = p_{\rm c} = 0$. Notice the transition is indeed critical, as exhibited by the asymptotic power law, 
 central curve in Figure 1.

	Being a multiplicative process, the natural choice for order parameter to describe 
 the transition is $-\log{w_{\rm max}}$, which vanishes in the collapsed phase, $p<0$. We compute this order 
 parameter through a time averaging procedure followed by a sample average. Each sample corresponds to an 
 initially random distribution of wealths, starting from which the above-defined dynamic rule is processed 
 during $T$ years. For each $n$, the time average of $\log{w_n}$ is then taken during the last $T/2$ years, 
 where $T$ is large enough to assure convergence (we adopted $T = 32{\rm M}$, large enough within our 
 numerical precision). The same process is repeated by generating $S$ different initial conditions  
 (we adopted $S = 8$). All time averages are then sample averaged, resulting in a final list 
 $<<-\log{w_n}>>$ with $N$ entries and the corresponding error bars. Figure 2 shows 
 one such distribution\footnote{All algebraic operations were performed within a $113$-bit mantissa, by 
 multiplying each fraction $w_n$ by $2^{60}$ and storing it in an integer variable with $60$ bits {\it plus} 
 an ordinary double precision variable with a $53$-bit mantissa. The resulting numerical accuracy on the 
 sum of wealths one needs in order to perform the annual renormalisation of the monetary unit is bounded 
 above $10^{-34}$ ($2^{-113}$). Were we to use simply a double precision variable, the accuracy would 
 drop to $10^{-15}$ ($2^{-53}$). That is why we restrict the plot to wealth shares larger than $10^{-34}$.}. 

	The order parameter is 

	$$m =\, <<-\log{w_1}>>\,\, , \eqno(1)$$

 \noindent the first entry of the quoted list. It is shown as a function of $p$ in Figure 3 (symbols at 
 the main plot), within the active phase $p>0$ (for $p<0$, the result is always $m = 0$ as it should be). 
 Near the transition point, the order parameter $m$ follows a power law $m \propto p^\beta$, with 
 $\beta \approx 0.83$ estimated by fitting Figure 3 (bottom right inset) for $p \le 0.01$. Compared with 
 the corresponding critical exponents $\beta$ normally observed in equilibrium phase transitions, this 
 value is unusually high. However, equilibrium thermodynamic concepts (like inequalities among critical 
 exponents) cannot be directly applied here. The traditional notation $\beta$ is used only by analogy.

	For completeness, it is interesting to provide some kind of external field acting on all agents 
 on the same footing. The field is expected to smooth the singularities at the transition point in the same 
 way an external magnetic field acting on a ferromagnetic system does. Redistribution plays the role 
 of such an external field $h$, defined as follows:

	$$\frac{1}{h} = -\log{R}\,\, . \eqno(2)$$

 \noindent Full redistribution corresponds to $h \to \infty$, whereas the no-redistribution limit 
 where the transition occurs corresponds\footnote{In principle, in order to avoid null wealths and keeping 
 the meaning of the system size $N$, one should set a very small $h>0$, similar to equilibrium phase 
 transitions where some residual external field is necessary in order to break the symmetry in finite 
 systems. When some particular wealth share $w_n$ becomes smaller than $10^{-318}$ during 
 the process, it is replaced by a copy of the previous one, $w_{n-1}$. Thus, the no-redistribution case 
 does not correspond exactly to $R=0$, but is equivalent to set $R \approx N \times 10^{-318}$.} 
 to $h \to 0$. Figure 3 shows smooth curves obtained for some values of $h$. Moreover, at the transition 
 point $p=0$, the order parameter $m$ follows another power law $m \propto h^{1/\delta}$, Figure 3 (upper 
 left inset). From it, one can estimate $\delta \approx 1.03$. Indeed, $\delta = 1$ is expected by the model 
 definition itself, since there are no interactions between agents. Only the annual redefinition of monetary 
 unit correlates agents to each other. This is different from interacting systems usually treated in 
 equilibrium phase transition studies, where the various elements behave collectively through some prescribed 
 interaction, thus displaying a non-linear response $m$ to the external stimulus $h$: $\delta$ would be larger 
 than unity in such a case.

\begin{figure}
\onefigure[width=\columnwidth]{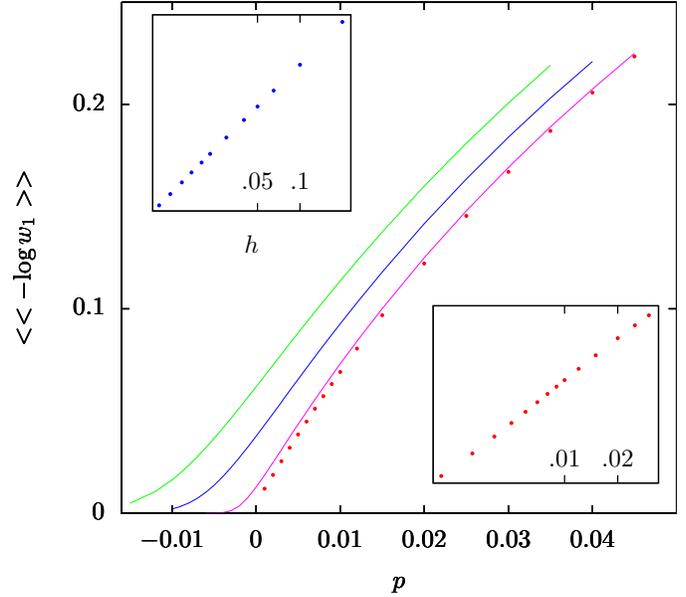}
\caption{Order parameter ($w_1$ refers to the richest agent) as a function of $p$. 
 Symbols at the main plot are obtained without redistribution, i.e. external field $h \to 0$, 
 exhibiting the characteristic singularity at the critical point $p=0$. Again, error bars are 
 smaller than the symbols. Smooth, continuous lines correspond to finite external fields 
 $h=0.05$, $h=0.03$ and $h=0.01$ from top down. Insets show the same quantity in 
 $log \times log$ plots as function of $p>0$ (without redistribution, bottom right) or $h$ 
 (at the critical point $p=0$, upper left): the straight lines confirm the critical character 
 of the transition. Here $N=10,000$, the same points are reproduced for $N=1,000$, within the 
 error bars.
}
\label{fig.3}
\end{figure}

\section{Deterministic, Mean Field Approximation}

	Let's restrict the treatment to the no-redistribution case where the transition occurs, 
 skipping step III. At time $t$ the fraction of agents whose wealth shares fall between 
 $w$ and $w+{\rm d}w$ is ${\rm d}w\, g_t(w)$, where

	$$\int_0^1 {\rm d}w\, g_t(w) = 1 \eqno(3)$$

 \noindent comes from agent counting normalisation, relative to the (large) fixed number $N$ of agents. 
 Another normalisation, 

	$$S = N <w>_t\, = N \int_0^1 {\rm d}w\, w\, g_t(w) = 1 \eqno(4)$$

 \noindent corresponds to the total wealth considered as monetary unit. Symbol $<\dots>_t$ indicates 
 the average over the current configuration, all agents, at time $t$. Instead of $g_t(w)$, let's define 
 the ranking function

	$$r_t(w) = N \int_w^1 {\rm d}w'\, g_t(w')\,\,\, , \eqno(5)$$

 \noindent which is the already quoted Zipf distribution of wealth shares (such as Figure 2, with 
 interchanged horizontal and vertical axes).	

	Consider $P(f)$ the arbitrary but fixed probability distribution for the multiplicative 
 factors $f_n$. After steps I, II and IV, the new ranking function is 

	$$r_{t+1}(w) = \int {\rm d}f\, P(f)\, r_t(x/f)\,\,\, , \eqno(6)$$

 \noindent where 

	$$x = \frac{(1-A)<f>}{2p}\, \Bigg(1-\sqrt{1-\frac{4pF_t\,w}{1-A}}\Bigg)\,\,\, ,\eqno(6a)$$

 \noindent $<\dots>$ means the average under the fixed probability distribution $P(f)$ and

	$$F_t = 1 - p\, \frac{<f^2>}{(1-A)<f>^2}\, \frac{<w^2>_t}{<w>_t}\,\,\, .\eqno(6b)$$

 \noindent The averages $<w>_t$ and $<w^2>_t$ can be determined directly from the ranking function, 
 instead of $g_t(w)$.

	The reasoning leading to equations (6) is based on solving backwards relation 
 $W'' = (1-A-pW'/S')W'$, within the mean field assumption $S' =\,\, <f>$, i.e. after step I the new 
 total wealth $S'$ is replaced by its average $<f>$ over all possible choices of multiplicative factors. 
 Then, we equate the fraction ${\rm d}W''\, \overline{\overline{g}}(W'')$ of agents whose wealths 
 are between $W''$ and $W'' + {\rm d}W''$ after taxes to the corresponding fraction 
 ${\rm d}W'\, \overline{g}(W')$ of agents whose wealths were between $W'$ and $W' + {\rm d}W'$ 
 before taxes.

	At this point, after applying equation (6), the up-to-now continuous ranking function should be 
 discretised along the $r$ axis in $N$ channels, before starting the next year. This procedure is necessary 
 in order to preserve the system size $N$, as already discussed.
 The plot $r \times w$ is divided in $N$ horizontal strips with equal heights, each strip then replaced by 
 a rectangle with the same area. Dynamic evolution, equation (6) plus discretisation, is then repeated 
 until convergence. 

	In practice, we divide the $w$ axis in $M$ channels ($M >> N$), assigning some tentative guess 
 for the final steady state ranking function $r(w)$ (any monotonically decreasing form between 
 $r(0)=N$ and $r(1)=0$). Taking a particular value $w$, equation (6) is applied and the 
 resulting $r$ replaces the original $r(w)$. The process is repeated for all channels, again and 
 again, until convergence (the so-called relaxation method). The result is as expected: $w_{\rm max} = 1$
 for $p < 0$.

	Returning to the simple case of a binary distribution of multiplicative factors ($f_n = 1$ or $2$ 
 with probabilities $50\%$), the very same mean field approach can be reformulated as follows.

 \vskip10pt a) Starting from the current wealth shares $w_n$ at time $t$, $n = 1, 2 \dots N$, one first 
 performs a copy of them, doubling all values of the copy.

 \vskip10pt b) Now, one has $2N$ wealths $W_n$ instead of the original $N$, summing to $3$ instead of unity. 
 Then, one applies taxes to these wealths.

 \vskip10pt c) After that, the resulting $2N$ wealths are listed in decreasing order. 

 \vskip10pt d) Finally, one restores the original population $N$: the average between the first and the 
 second largest wealths is assigned to the first agent, the average between the third and fourth wealths 
 is assigned to the second agent, so on.

 \vskip10pt A comment about this new procedure follows. The averaging step (d) highlights the mean-field 
 character of the current deterministic approach, compared with the stochastic formulation where each  
 wealth can be doubled or not. In the stochastic version, at the critical point, 
 $1 - w_{\rm max}$ slowly vanishes as time goes by as shown in Figure 1, central plot, whereas it 
 converges to $1/2$ in the mean-field approximation generating a first-order transition 
 gap\footnote{Indeed, for $p=0$ the whole list of wealth shares converges to $w_n = 0.5^n$.}.
 The transition point $p = 0$, however, remains the same. 

	Another comment. This simple alternative formulation bypasses the ranking function; its role is 
 automatically performed by the ordering step (c) applied to the set of $2N$ wealths.

	Still another comment. One needs only two copies of the wealth shares, step (a), in the particular case 
 when one adopts only two possible wealth multiplicative factors, here $1$ or $2$. In general, one needs one 
 copy for each possible wealth multiplicative factor $f_n$, taking into account their probability distribution 
 $P(f)$. In this case, the original formulation by relaxing the ranking function may be more economical 
 for computer calculations. Anyway, the two approaches are completely equivalent.

\section{Conclusions}

	The wealth distribution of a population is submitted to a multiplicative process followed by 
 regressive or progressive taxation, where tax rates are higher for poor than rich agents or vice-versa,   
 respectively. In the long term, regressive taxation leads to social collapse, all 
 wealth falls in the hands of a single agent, whereas it remains forever distributed among agents if 
 progressive taxation is adopted instead. This transition is different from that studied in previous works, 
 for instance \cite{Bouchaud}, in which the so-called wealth condensation is the main issue, the 
 concentration of the entire wealth in the hands of a {\sl finite} number of agents within an 
 otherwise infinite population. In some sense, without redistribution, our model always eventually 
 leads to such condensation, even in the active phase. In this case, the wealths listed in decreasing 
 order follow an almost exponential decay, Figure 2, configuring an {\sl effective} finite number 
 of agents participating in the process (inversely proportional to the slope of plots 
 like Figure 2). One way to avoid such condensation is the uniform redistribution of the collected taxes, 
 as in \cite{Bouchaud}, at the end of each year. Doing so, uniformly redistributing a fraction $R$ 
 of the collected taxes, the transition singularities disappear in the same way external fields do in 
 the traditional theory of equilibrium phase transitions (curves become smooth, as shown in Figure 3). 

	As a final comment, we notice that progressive taxation ($p>0$) alone is not able to solve the 
 inequality problem, since only a finite number of agents effectively participate of the wealth exchange 
 game. Even for progressive taxation, wealth concentration still holds, not in the hands of a single agent but 
 in the hands of a finite number. In order to avoid wealth concentration, economic policy makers should 
 not only adopt progressive taxation, but also implement some kind of redistribution. Just after 
 World War II, policy makers perceived this fact, creating a set of social programs 
 now known as the ``welfare state''.

\acknowledgments

	I am indebted to Evaldo Curado and Cristian Moukarzel for helpful discussions, and 
 to Ronald Dickman for a critical reading of the manuscipt.

\end{document}